\documentclass[aps,prl,twocolumn,showpacs,groupedaddress]{revtex4}  
\usepackage{graphicx}  
\usepackage{dcolumn}   
\usepackage{bm}        
\usepackage{amssymb}   
\usepackage{amsmath}
\usepackage{epstopdf}
\usepackage{ragged2e}

\newcommand {\be}{\begin{eqnarray}}
\newcommand {\ee}{\end{eqnarray}}  
\newcommand {\ba}{\be \begin{aligned}} 
\newcommand {\ea}{\end{aligned} \ee}
\newcommand {\bx}{\mathbf{x}}
\newcommand {\bk}{\mathbf{k}}
\newcommand {\bq}{\mathbf{q}}

\newcommand {\br}{\mathbf{r}}
\newcommand {\by}{\mathbf{y}}

\begin{document}

\title{Majorana bands, Berry curvature, and thermal Hall conductivity in the vortex state of a chiral p-wave superconductor}
\author{James M. Murray}
\affiliation{National High Magnetic Field Lab and Department of Physics, \\
	Florida State University, Tallahassee 32310, USA}
\author{Oskar Vafek}
\affiliation{National High Magnetic Field Lab and Department of Physics, \\
	Florida State University, Tallahassee 32310, USA}
\date{\today}

\begin{abstract}
\noindent Majorana quasiparticles localized in vortex cores of a chiral p-wave superconductor hybridize with one another to form bands in a vortex lattice. We begin by solving a fully microscopic theory describing all quasiparticle bands in a chiral p-wave superconductor in magnetic field, then use this solution to build localized Wannier wavefunctions corresponding to Majorana quasiparticles. A low-energy tight-binding theory describing the intervortex hopping of these is then derived, and its topological properties---which depend crucially on the signs of the imaginary intervortex hopping parameters---are studied. We show that the energy gap between the Majorana bands may be either topologically trivial or nontrivial, depending on whether the Chern number contributions from the Majorana bands and those from the background superconducting condensate add constructively or destructively. This topology directly affects the temperature-dependent thermal Hall conductivity, which we also calculate. 
\end{abstract}

\maketitle

It has long been known that an isolated vortex core in a chiral p-wave superconductor can host zero-energy Majorana modes \cite{volovik99,read00,beenakker13,elliott15}. Such topological superconductivity may be present in Sr$_2$RuO$_4$ \cite{maeno11,kallin12}, and also has analogues in the Moore-Read fractional quatum Hall state \cite{moore91,read00} and in the A-phase of superfluid $^3$He \cite{vollhardt13}. 
When multiple vortices are brought close together, the Majorana zero modes hybridize, leading to states at positive and negative energy. From studying pairs of vortices, it has been shown that the sign as well of the amplitude of this energy splitting depends on the intervortex distance \cite{cheng09,cheng10}. 

Just as there are two possible signs of the energy splitting for a pair of vortices, there are two topologically distinct possibilities for the energy gap between the Majorana bands in a system consisting of many vortices \cite{ludwig11,lahtinen12,lahtinen14}. In this work we provide a way in which to compute the topological properties of the ground state starting from a microscopic Hamiltonian describing paired electrons on a square lattice. We show how the topologically trivial and nontrivial states can be understood as arising due to destructive or constructive addition of the Chern number contributions from the Majorana bands and  from the background superconducting condensate, respectively. We further show that these different topological states can lead to different behaviors of the intrinsic contribution to the thermal Hall conductivity, which we compute explicitly from the microscopic model.

Several authors have recently studied periodic arrays of vortices starting from a continuum model of a chiral p-wave superconductor \cite{biswas13,silaev13,zhou14}. 
Unlike Ref.~\cite{biswas13}, we do not find any zero-energy flat band with zero Chern number. This disagreement arises due to the neglect of the magnetic field in that work, as well as the different choice of the spatial profile of the complex phase of the superconducting order parameter.
Our work differs from Ref.~\cite{zhou14} in that we do not find any strong anisotropy emerging in our low-energy theory describing Majorana quasiparticles. Further, the authors of that work find a single edge mode in their low-energy theory, whereas an even number always emerges from our theory. At present, we do not understand the precise reason for this disagreement.
Our results extend those of the semiclassical calculation in Ref.~\cite{silaev13} by providing a fully quantum treatment, which is necessary in order to address effects associated with Berry phases and topological properties.

Finally, very recently another work that has some overlap with our results has appeared \cite{liu15}. As in our work, the authors begin from a lattice Hamiltonian describing a chiral p-wave superconductor in magnetic field and solve for the full quasiparticle bandstructure. They then obtain a tight-binding description of the Majorana bands by fitting the hopping parameters to this full solution. This is a different procedure from our work, in which {\em deriving} a tight-binding model from localized Wannier functions allows us to determine the signs of these hopping parameters rather than just the absolute values. As we will show below, these signs are what determine the topological properties of the Majorana bands, and hence have a direct affect on the Berry curvature and thermal Hall conductivity. The calculation of these quantities is the main result of our work and did not appear in Ref.~\cite{liu15}.

\section{Microscopic lattice model}
\label{sec:model}
Following the approach of Refs.~\cite{vafek01a,cvetkovic15}, we use the following Hamiltonian to describe a superconductor on a square lattice in a magnetic field:
\begin{widetext}
\be
\label{eq:0428a}
\mathcal{H}=\sum_{{\bf r}}\left(\sum_{\delta=\hat{x},\hat{y}}
	\bigg[ t_{{\bf r},{\bf r}+\delta}
	c^{\dagger}_{{\bf r},\sigma}c_{{\bf r}+\delta,\sigma}
	+\Delta_{{\bf r},{\bf r}+\delta}
	\left(c^{\dagger}_{{\bf r},\uparrow}c^{\dagger}_{{\bf r}+\delta,\downarrow}
	-c^{\dagger}_{{\bf r},\downarrow}c^{\dagger}_{{\bf r}+\delta,\uparrow}\right)
	+H.c.\bigg]
	-\mu_\sigma c^{\dagger}_{{\bf r},\sigma}c_{{\bf r},\sigma} \right).
\ee
\end{widetext}
Here $c_{\br,\sigma}$ is the electron operator, with implicit summation over spin $\sigma = \uparrow,\downarrow$. The hopping, which includes a Peierls phase factor due to the magnetic field, is given by $t_{\br,\br+\delta} = -t e^{-i A_{\br,\br+\delta}}$, where $A_{\br,\br+\hat\bx} = \pi y \Phi/\phi_0$, $A_{\br,\br+\hat\by} = -\pi x \Phi/\phi_0$, $\phi_0 = hc/e$ is the elementary flux quantum, and $\Phi$ is the magnetic flux through each plaquette. The superconducting pairing term is $\Delta_{\br, \br+\delta} = \Delta_\delta e^{i\theta(\br)}\exp{\left( \frac{i}{2} \int_\br^{\br+\delta} d\mathbf{l}\cdot \nabla \theta \right)}$. For chiral p-wave pairing, the amplitudes are given by $(\Delta_{\br,\br\pm\hat\bx}, \Delta_{\br,\br\pm\hat\by}) = (\pm \Delta, \pm i \Delta)$ or $(\pm \Delta, \mp i \Delta)$ for ``$p_x +i p_y$'' or ``$p_x -i p_y$'' pairing, respectively. As discussed in previous works \cite{vafek01a,cvetkovic15}, the phase of the superconducting order parameter $\theta(\br)$ is chosen to be a solution to the continuum London equations:
\ba
\nabla\times\nabla \theta(\br) &= 2\pi \hat z \sum_j \delta (\br - \br_j), \\
\nabla^2 \theta(\br) &= 0.
\ea
Finally, the chemical potential term in general includes a spin-dependent Zeeman term: $\mu_{\uparrow,\downarrow} = \mu \pm h_Z$. When written in the Bogoliubov-de Gennes language, the Zeeman term is proportional to the identity matrix in the Hamiltonian (see \eqref{eq:0428b} below), so that $h_Z$ merely shifts the energy bands as a chemical potential would for an ordinary fermionic system without superconductivity. We consider the effects of this shift when calculating the thermal Hall conductivity below.

\begin{figure}
\centering
\includegraphics[width=0.3\textwidth]{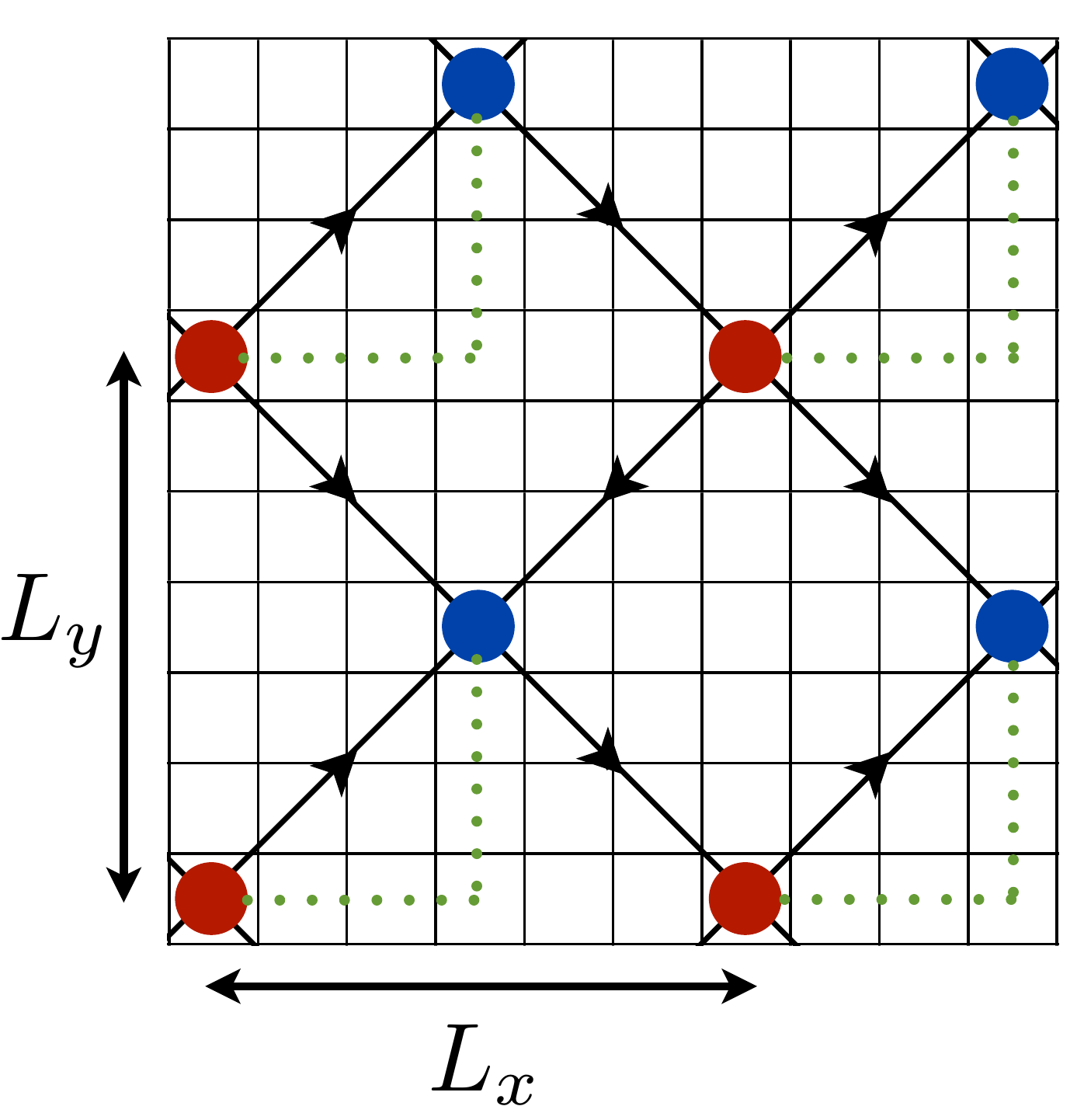}
\caption{Square vortex lattice, with arrows indicating the sign of the imaginary nearest-neighbor hopping of fermionic quasiparticles between vortices. The dotted lines indicate branch cuts connecting the two vortices within each magnetic unit cell.
\label{fig:0504a}}
\end{figure}

We consider a magnetic unit cell of $L_x \times L_y$ sites, with each unit cell containing two vortices, as shown in Figure \ref{fig:0504a}. 
The superconductor is assumed to be strongly type-II, with a sufficiently small vortex core size and suffiently large penetration depth that the amplitudes of both the order parameter and magnetic field are constant in space. By employing a singular gauge transformation \cite{franz00,vafek01a,vafek06,cvetkovic15}, the Hamiltonian \eqref{eq:0428a} can be made periodic with spatial period $L_{x(y)}$ along the $x$ $(y)$ direction, thereby enabling the use of Bloch's theorem. Performing the following particle-hole transformation: 
\be
\label{eq:0428b}
   \left(
   \begin{matrix} 
      c_{\br,\uparrow} \\
      c^\dagger_{\br,\downarrow} 
   \end{matrix}
   \right)
    = \frac{1}{\sqrt{N_\mathrm{uc}}}\sum_\bk e^{i\bk\cdot\br}
   \left(
   \begin{matrix} 
      e^{i\theta(\br)/2}\psi_{\br,\uparrow} (\bk) \\
      e^{-i\theta(\br)/2}\psi_{\br,\downarrow} (\bk)
      \end{matrix}
   \right),
\ee
where $k_{x,y} \in (\frac{-\pi}{L_{x,y}}, \frac{\pi}{L_{x,y}}]$, and the new fermionic operators satisfy $\psi_{\br+\mathbf{R},\sigma} (\bk) = \psi_{\br,\sigma} (\bk)$, the Hamiltonian \eqref{eq:0428a} becomes
\begin{widetext}
\ba
\label{eq:0428c}
\mathcal{H}=
	&\sum_{{\bf r}\in u.c.}\sum_{{\bf k}}\left(\sum_{\delta=\hat{x},\hat{y}}\left[
	e^{i{\bf k}\cdot\delta}\left(t^{\uparrow\uparrow}_{{\bf r},{\bf r}+\delta}\psi^{\dagger}_{{\bf r},\uparrow}(\bk)\psi_{{\bf r}+\delta,\uparrow}(\bk)
- t^{\downarrow\downarrow}_{{\bf r},{\bf r}+\delta}\psi^{\dagger}_{{\bf r},\downarrow}(\bk)\psi_{{\bf r}+\delta,\downarrow}(\bk)\right)+H.c.\right]
	-\tilde\mu_{\sigma}\psi^{\dagger}_{{\bf r},\sigma}(\bk)
	\psi_{{\bf r},\sigma}(\bk)\right) \\
&+ \sum_{{\bf r}\in u.c.}\sum_{{\bf k}}\sum_{\delta=\hat{x},\hat{y}}\left[\lambda^{\uparrow\downarrow}_{{\bf r},{\bf r}+\delta}
\left(e^{i{\bf k}\cdot\delta}
\psi^{\dagger}_{{\bf r},\uparrow}(\bk)\psi_{{\bf r}+\delta,\downarrow}(\bk)+
e^{-i{\bf k}\cdot\delta}
\psi^{\dagger}_{{\bf r}+\delta,\uparrow}(\bk)\psi_{{\bf r},\downarrow}(\bk)
\right)+H.c.\right] \\
\equiv & \sum_\bk \sum_{\br,\br' \in u.c.} \sum_{\sigma, \sigma'}
	\psi^\dagger_{\br,\sigma}(\bk) \mathcal{H}_\bk (\br, \sigma; \br', \sigma')
	\psi_{\br',\sigma'}(\bk),
\ea
\end{widetext}
where $t^{\uparrow\uparrow}_{{\bf r},{\bf r}+\delta}={t^{\downarrow\downarrow}}^*_{{\bf r},{\bf r}+\delta} =
t_{{\bf r},{\bf r}+\delta}e^{\frac{i}{2}\theta({{\bf r}+\delta})}e^{-\frac{i}{2}\theta({\bf r})}$, $\lambda^{\uparrow\downarrow}_{{\bf r},{\bf r}+\delta}=\Delta_{{\bf r},{\bf r}+\delta}e^{-\frac{i}{2}\theta({{\bf r}+\delta})}e^{-\frac{i}{2}\theta({{\bf r}})}$, and
$\tilde\mu_{\uparrow(\downarrow)}=\pm\mu + h_Z$. 
The branch cuts in $e^{i\theta(\br)/2}$ are chosen to connect vortices pairwise within a unit cell, as shown in Figure \ref{fig:0504a}. (Further details regarding the treatment of branch cuts can be found in Ref.~\cite{cvetkovic15}.)

\begin{figure}
\centering
\includegraphics[width=0.48\textwidth]{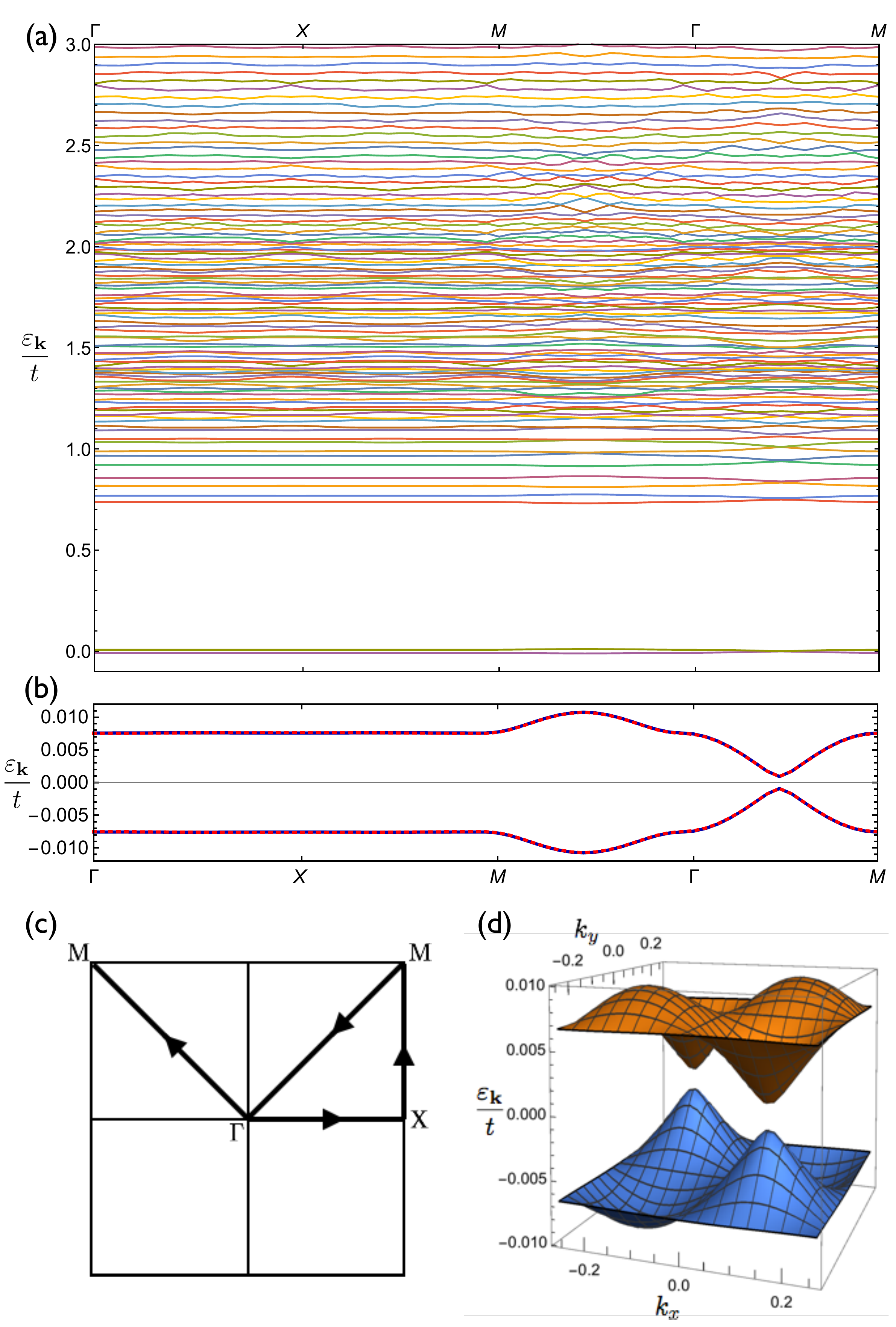}
\caption{(a) Quasiparticle energy bands of a p-wave superconductor with $p_x-ip_y$ pairing in a perpendicular magnetic field, for a square vortex lattice with intervortex distance  $L_x = L_y = 12$, pairing strength $\Delta/t = 0.5$, chemical potential $\mu / t = -2$, and Zeeman shift $h_Z = 0$. (b) Detailed view of the two Majorana bands, where the dashed lines are obtained by solving the full Hamiltonian \eqref{eq:0428c}, while the (virtually indistinguishable) solid lines come from the effective tight-binding theory with first- and second-neighbor hoppings. (c) Magnetic Brilluoin zone, showing the path along which the energy bands are plotted. (d) The Majorana bands feature two Dirac cones along the line $k_y = -k_x$, which are gapped due to next-nearest neighbor hopping.
\label{fig:0616a}}
\end{figure}

Figure \ref{fig:0616a} shows the band structure obtained by diagonalizing the Hamiltonian \eqref{eq:0428c}.
The spectrum features a large gap with energy $\sim \Delta$, with two bands near $\varepsilon = 0$ arising from the intervortex tunneling of Majorana zero modes. The bandwidth of these low-energy bands decreases exponentially with the distance between vortices. Upon closer inspection, one finds that the Majorana bands feature two gapped Dirac cones, which is expected from the fact that the imaginary hopping parameters describe a $\pi$-flux model, as pointed out previously \cite{grosfeld06} and described in detail below. 

\begin{figure}
\centering
\includegraphics[width=0.4\textwidth]{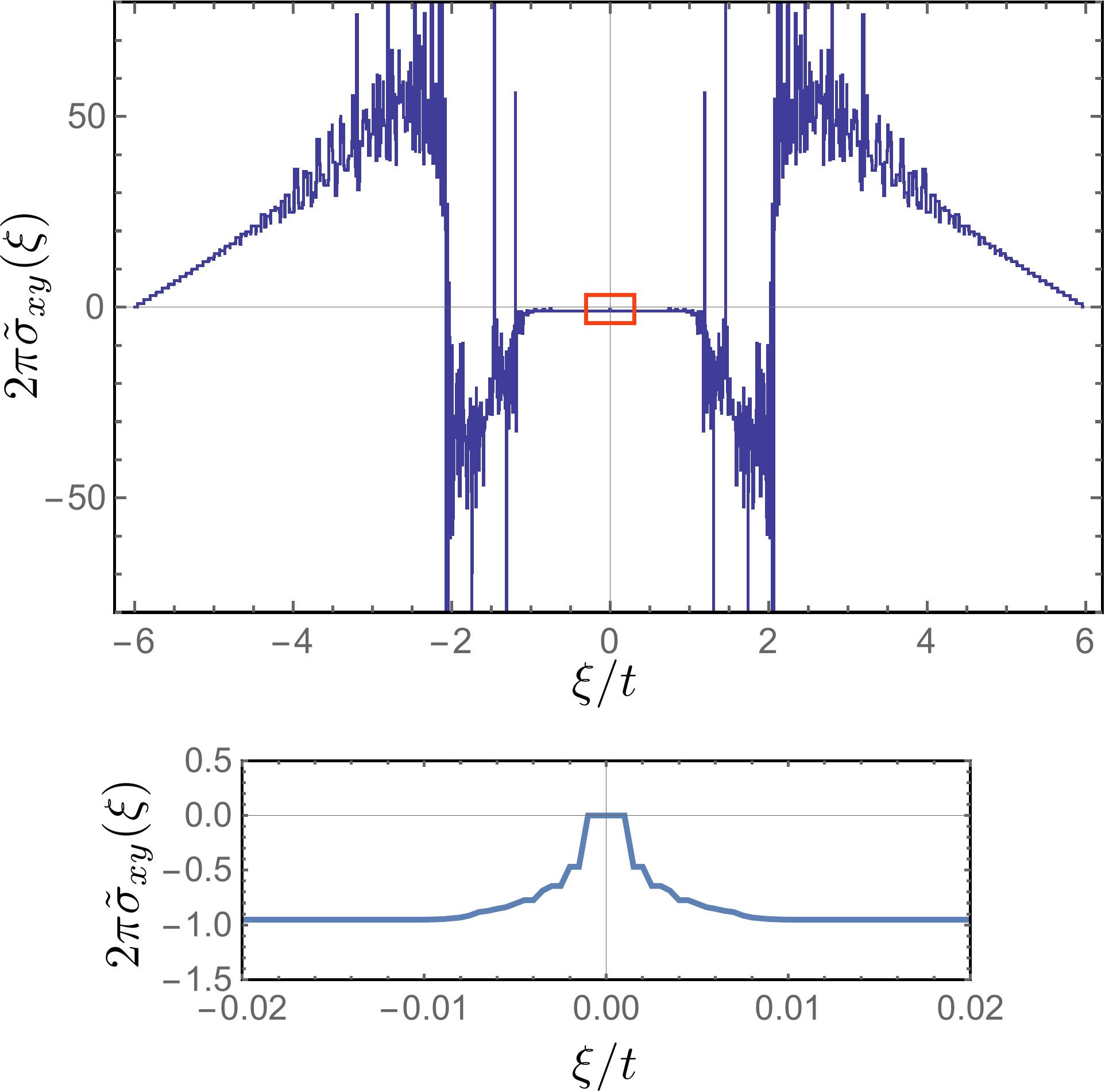}
\caption{The integrated Berry curvature, summed over bands up to energy $\xi$, with same parameters as in Figure \ref{fig:0616a}. The lower panel provides a detailed view near $\xi=0$, showing that the Chern number within the large energy gap is $\nu=-1$, while the Majorana bands carry Chern numbers $\nu = \pm 1$.
\label{fig:0620a}}
\end{figure}

Using the numerically determined eigenstates of the Hamiltonian \eqref{eq:0428c}, the integrated Berry curvature summed over all occupied bands is given by \cite{smrcka77,thouless82,cvetkovic15}
\ba
\label{eq:0609b}
&\tilde\sigma_{xy} (\xi) = -i \int \frac{d^2 k}{(2\pi)^2} 
	\sum_{E_m(\bk) < \xi < E_n(\bk)} \\
&\quad\times
	\frac{\bigg\langle m\bk \left| \frac{\partial\mathcal{H}_\bk}{\partial k_x}
	\right| n\bk \bigg\rangle
	\bigg\langle n\bk \left| \frac{\partial\mathcal{H}_\bk}{\partial k_y}
	\right| m\bk \bigg\rangle - (x \leftrightarrow y)}{(E_m(\bk) - E_n(\bk))^2}.
\ea
In this equation, $m$ and $n$ denote the quasiparticle bands, and the summation is over all $m$ and $n$ subject to the constraint $E_m(\bk) < \xi < E_n(\bk)$. Methods for evaluating this expression efficiently were discussed in Ref.~\cite{cvetkovic15}. Figure \ref{fig:0620a} shows the behavior of $\tilde \sigma_{xy} (\xi)$. 
Near the top and bottom portions of the energy spectrum, the quasiparticle bands are either electron- or hole-like, so that $\tilde \sigma_{xy} (\xi)$ just sums up the Chern numbers $\pm 1$ as $\xi$ passes through each quasiparticle band. A large jump from positive to negative values occurs at the van Hove singularity, where there is a pile-up of Berry curvature. For $|\xi| \lesssim \Delta$, $\tilde \sigma_{xy} (\xi)$ is suppressed due to the fact that the quasiparticle states are superpositions of electron and hole states, and hence carry little or no Berry curvature. Closer inspection, though, reveals an interesting structure in this region. Rather than completely vanishing, the integrated Berry curvature within the large energy gap takes a value $\pm 1$ for $p_x\pm ip_y$ pairing, reflecting the topological nature of the superconducting state. At the lowest energies, however, the Chern number becomes either 0 or $\pm 2$. As we shall show below, this reflects the fact that the Majorana bands carry Chern number $\pm 1$. One thus sees that the topological nature of the gap between the Majorana bands depends on the interplay of the topologies of the Majorana bands and of the background condensate.

\section{Effective tight-binding model for Majorana fermions}

In order to better understand the topological nature of the Majorana bands, we proceed to derive a $2 \times 2$ tight-binding Hamiltonian describing the hopping of Majorana quasiparticles between vortices. Given Bloch eigenstates $\psi_{n\bk}(\br,\sigma)$ of the Hamiltonian \eqref{eq:0428c}, Wannier functions can be defined as
\be
\label{eq:0503a}
w_{\alpha} (\br - \mathbf{R} - \br_\alpha, \sigma) 
	= \frac{1}{N_\mathrm{uc}}\sum_{n\bk} U_{n \alpha}^{(\bk)} e^{-i\bk\cdot (\mathbf{R} + \br_\alpha)}
	\psi_{n\bk}(\br,\sigma),
\ee
where $n$ is a quasiparticle band index, $N_\mathrm{uc}$ is the number of points in momentum space, and $\alpha = A,B$ denotes the two vortex sublattices. The vectors  $\mathbf{R}$ and $\br_\alpha$ point to a magnetic unit cell and to the position of a vortex of type $\alpha$ within the unit cell, respectively. In building a low-energy theory, we keep only the values of $n$ corresponding to the two Majorana bands. In this case $U_{\alpha n}^{(\bk)}$ is an arbitrary $2\times 2$ unitary matrix at each $\bk$, which we want to choose such that each Wannier function is localized at one of the vortex cores located at $\br = \mathbf{R} + \br_\alpha$. This can be accomplished straightforwardly using projection operators starting from a trial wavefunction \cite{deCloizeaux64,marzari12}. The first step is to define initial guesses $|g_{\mathbf{R}\alpha} \rangle$ for the Wannier functions, which we take to be real Gaussian functions localized near each vortex core. We then project these trial wavefunctions onto the Bloch functions for the two Majorana bands:
\be
\label{eq:0503b}
|\phi_{\alpha\bk}\rangle = 
	\sum_m |\psi_{m\bk}\rangle \langle\psi_{m\bk} | g_{\mathbf{R}=0,\alpha}\rangle .
\ee
Unlike the original Bloch functions, which have an arbitrary phase at each $\bk$, these states will have phases that are smooth as a function of $\bk$. The next step is to orthonormalize these states to obtain pseudo-Bloch functions:
\be
\label{eq:0503c}
|\tilde\psi_{\alpha\bk}\rangle 
	= \sum_n |\phi_{\alpha\bk}\rangle (S_\bk^{-1/2})_{n\alpha},
\ee
where $(S_\bk)_{n\alpha} = \int_{u.c.} \langle \phi_{n\bk} | \phi_{\alpha\bk}\rangle$ (the integral refers to averaging over the unit cell). The Wannier functions are then given by
\be
\label{eq:0503d}
|w_{\mathbf{R}\alpha}\rangle 
	= \frac{1}{N_\mathrm{uc}} \sum_\bk e^{-i\bk\cdot (\mathbf{R} + \br_\alpha)} 
	|\tilde\psi_{\alpha\bk}\rangle.
\ee
The coefficients appearing in \eqref{eq:0503a} are thus given by
\be
\label{eq:0503e}
U_{n \alpha}^{(\bk)} = \langle\psi_{n\bk} | \tilde\psi_{\alpha\bk}\rangle
	= \sum_{\br\sigma} \psi^*_{n\bk}(\br,\sigma) 
	\tilde\psi_{\alpha\bk} (\br,\sigma).
\ee

With the localized Wannier functions in hand, we are now in a position to derive a low-energy tight-binding Hamiltonian to describe the hopping of quasiparticles between vortex sites. We begin by expanding the fermion operators appearing in \eqref{eq:0428c} (putting a hat on the operators to avoid confusion) in a basis of Bloch eigenstates:
\be
\label{eq:0503f}
\hat\psi_{\br\sigma} 
\equiv e^{-i\bq\cdot\br} \hat\psi_{\br,\sigma}(\bq)
	= \sum_{\bk n} \psi_{\bk n}(\br, \sigma) \hat a_n(\bk).
\ee
We can then invert \eqref{eq:0503a} to obtain the Bloch states in terms of Wannier functions:
\be
\label{eq:0503g}
\psi_{n\bk} (\br,\sigma)
	= \sum_{\mathbf{R}\alpha} (U_{\alpha n}^{(\bk)})^*
	e^{i\bk\cdot (\mathbf{R} + \br_\alpha)}
	w_\alpha (\br - \mathbf{R} - \br_\alpha, \sigma) ,
\ee
and rewrite the Hamiltonian as
\ba
\label{eq:0503h}
& \sum_{\br\sigma}\sum_{\br'\sigma'} \hat\psi_{\br\sigma}^\dagger 
	\mathcal{H}(\br\sigma,\br'\sigma') \hat\psi_{\br'\sigma'} \\
& \quad =	\sum_{\br\sigma}\sum_{\br'\sigma'}\sum_{n\bk}\sum_{n'\bk}
	\hat a_n^\dagger(\bk) \hat a_{n'}(\bk') \\
&\quad\quad\quad \times \psi^*_{n\bk}(\br,\sigma) \mathcal{H}(\br\sigma,\br'\sigma') 
	\psi_{n'\bk'}(\br',\sigma'),
\ea
where $\mathcal{H}(\br\sigma,\br'\sigma') \equiv \mathcal{H}_{\bk=0}(\br\sigma,\br'\sigma')$ is the Hamiltonian in real space before applying Bloch's theorem.
Using \eqref{eq:0503g} and defining the new fermion operators that create quasiparticles on vortex sublattice $\alpha$
\be
\label{eq:0503i}
\hat d_\alpha^\dagger(\bk) = \sum_n U_{n\alpha}^{(\bk)} \hat a_n^\dagger(\bk),
\ee
the Hamiltonian becomes
\be
\label{eq:0503j}
\mathcal{H} = \sum_{\alpha\alpha'} \sum_\bk \hat d_\alpha^\dagger(\bk)
	\mathcal{H}_{\alpha\alpha'}(\bk) \hat d_{\alpha'}(\bk),
\ee
where
\ba
\label{eq:0503k}
& \mathcal{H}_{\alpha\alpha'}(\bk) = \sum_{\br\sigma}\sum_{\br'\sigma'} 
	\sum_{\Delta\mathbf{R}} 
	e^{i\bk\cdot(\Delta\mathbf{R} + \br_{\alpha'} - \br_\alpha)} \\
& \times w^*_\alpha (\br - \br_\alpha, \sigma) \mathcal{H}(\br\sigma,\br'\sigma')
	w_{\alpha'}(\br' - \Delta\mathbf{R} - \br_{\alpha'}, \sigma').
\ea
Due to the localization of the Wannier functions, only a limited number of terms with small $\Delta\mathbf{R}$ will give significant contributions to \eqref{eq:0503k}.

While the sums in \eqref{eq:0503k} could be computed using the Wannier functions constructed in \eqref{eq:0503d}, a simpler approach is to reexpress these Wannier functions in terms of the Bloch function eigenstates of $\mathcal{H}_\bk (\br\sigma,\br'\sigma') = e^{-i\bk\cdot\br} \mathcal{H} (\br\sigma,\br'\sigma') e^{i\bk\cdot\br'}$. One thus obtains
\begin{widetext}
\ba
\label{eq:0503l}
\mathcal{H}_{\alpha\alpha'}(\bk) &= \sum_{\br\sigma}\sum_{\Delta\mathbf{R}} 
	\sum_{\bq n} \sum_{\bq' n'}
	e^{i\bk\cdot(\Delta\mathbf{R} + \br_{\alpha'} - \br_\alpha)}
	e^{i \bq\cdot\br_\alpha} e^{-i \bq'\cdot(\Delta\mathbf{R} + \br_{\alpha'})}
	(U_{n \alpha}^{(\bq)})^* U_{n' \alpha'}^{(\bq')}
	\psi_{n\bq}^*(\br,\sigma) E_{n'\bq'} \psi_{n'\bq'}(\br,\sigma) \\
&= \sum_{\Delta\mathbf{R}}
	e^{i\bk\cdot(\Delta\mathbf{R} + \br_{\alpha'} - \br_\alpha)}
	\sum_{\bq n} e^{-i\bq\cdot(\Delta\mathbf{R} + \br_{\alpha'} - \br_\alpha)}
	(U_{n \alpha}^{(\bq)})^* U_{n \alpha'}^{(\bq)} E_{n\bq}.
\ea
\end{widetext}
Given the $U_{\alpha n}^{(\bq)}$ determined by the Wannier functions in \eqref{eq:0503e} and the energies for the Majorana bands shown in Figure \ref{fig:0616a}, this sum can be easily evaluated numerically to determine the tight-binding parameters describing hopping between vortex cores separated by distance $\Delta\mathbf{R} + \br_{\alpha'} - \br_\alpha$. 


For a square vortex lattice, the nearest- and next-nearest-neighbor hoppings are all imaginary, with signs of the nearest-neighbor hoppings as shown in Figure \ref{fig:0504a}.
This corresponds to a $\pi$-flux state, in which each plaquette is penetrated by a half quantum of magnetic flux \cite{affleck88,kotliar88,wen04,grosfeld06}. The low-energy effective Hamiltonian takes the form
\ba
\label{eq:0505a}
\mathcal{H}_M(\bk) =& 
	-t_0 \sin \left(\tfrac{k_x L_x + k_y L_y}{2}\right) \sigma_1 \\
	& + t_0 \cos \left(\tfrac{k_x L_x - k_y L_y}{2}\right) \sigma_2 \\
	& + t_1 [ \sin (k_x L_x) -  \sin(k_y L_y)] \sigma_3 .
\ea
The real parameters $t_0$ and $t_1$ correspond to nearest- and next-nearest-neighbor hopping, respectively \footnote{The tight-binding parameters determined from \eqref{eq:0503l} also have small real parts, the amplitudes of which are at least two orders of magnitude smaller than the imaginary hoppings. We take these to be spurious contributions due to the imperfect localization of the numerically determined Wannier functions and do not consider them further.}. Diagonalizing \eqref{eq:0505a} leads to energy bands with dispersion
\ba
\label{eq:0503n}
\varepsilon_\bk^{\pm} = \pm \bigg[ &
	t_0^2 \cos^2 \left(\tfrac{k_x L_x - k_y L_y}{2}\right)
	+ t_0^2\sin^2 \left(\tfrac{k_x L_x + k_y L_y}{2}\right) \\
	&+ t_1^2 [ \sin (k_x L_x) - \sin (k_y L_y) ]^2 \bigg]^{1/2},
\ea
which describes two massless Dirac cones at the points $(k_x, k_y) = \pm (\pi/L_x, -\pi/L_y)$ when $t_1 = 0$, with these cones becoming gapped when $t_1 \neq 0$, as shown in Figure \ref{fig:0616a} \footnote{As pointed out previously \cite{liu15}, different choices of Z$_2$ gauges in the low-energy Hamiltonian describing Majorana quasiparticles may cause the Dirac cones to appear at different locations in the magnetic Brilluoin zone.}.

\begin{figure}
\centering
\includegraphics[width=0.48\textwidth]{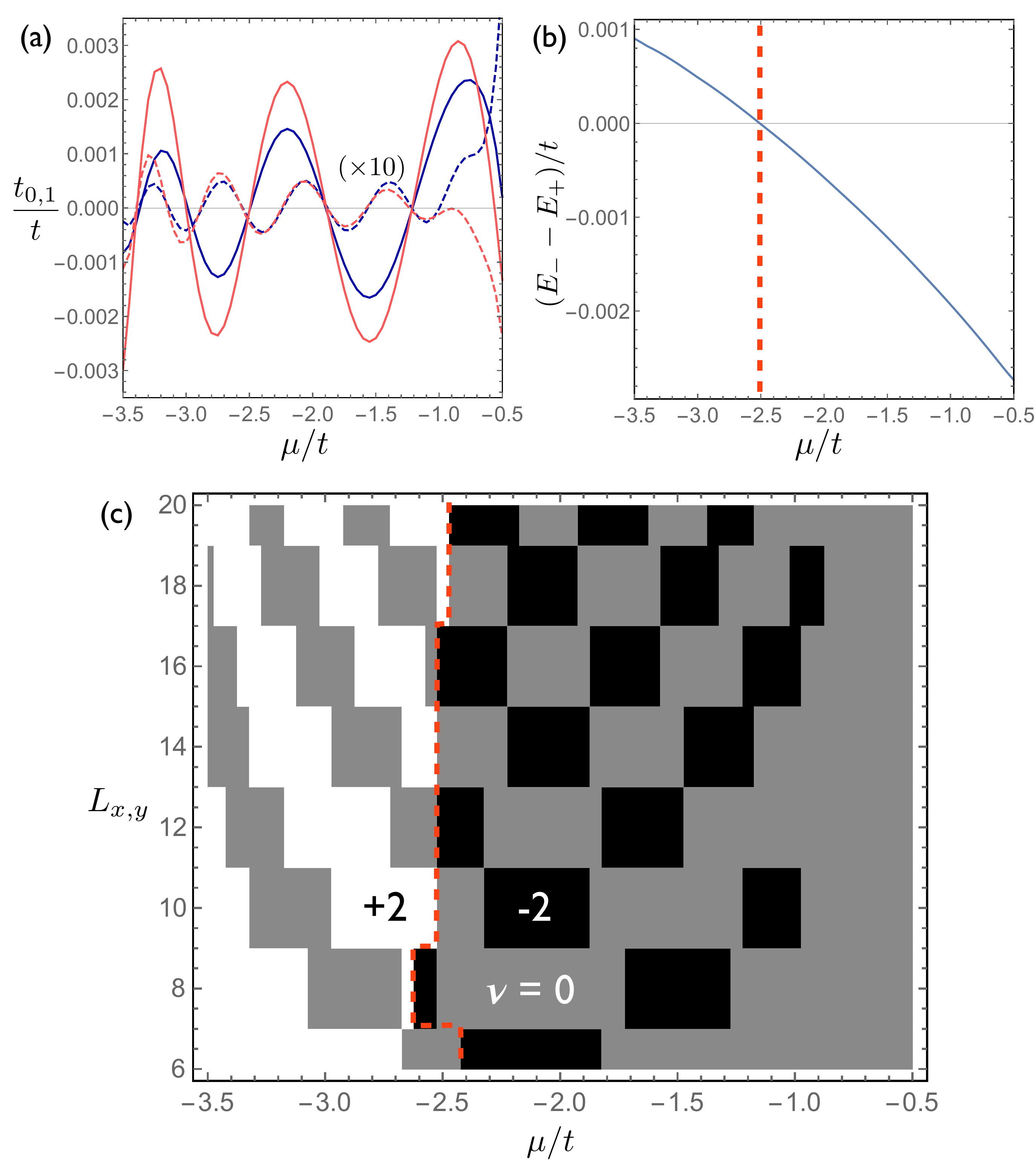}
\caption{(a) Oscillations in the nearest- and next-nearest-neighbor hopping parameters for the Majorana bands described by the low-energy effective Hamiltonian \eqref{eq:0505a}. Solid (dashed) lines correspond to $t_0$ ($t_1$, multiplied $\times 10$), and blue (red) lines correspond to $p_x-ip_y$ ($p_x+ip_y$) pairing, with magnetic length $L_x=L_y=16$, chemical potential $\mu/t = -2$, and pairing strength $\Delta/t = 0.5$. (b) Energy difference per site between the two pairing states, showing that $p_x-ip_y$ pairing is favored for larger values of $\mu$. (c) Phase diagram showing the total Chern number in the gap between Majorana bands, which is determined by the type of pairing and the sign of $t_1$. The pairing is $p_x+ip_y$ to the left of the red line and $p_x-ip_y$ to the right.
\label{fig:0616c}}
\end{figure}

As expected from previous studies \cite{cheng09,cheng10,liu15}, the tight-binding parameters $t_0$ and $t_1$ oscillate as a function of $k_F L$, where $k_F$ is the Fermi wavevector and $L = \sqrt{L_x L_y}$ is the magnetic length. Because $k_F$ is determined by the chemical potential, we show the oscillations of $t_0$ and $t_1$ as a function of $\mu$ in Figure \ref{fig:0616c}.
We also find, as shown in Figure \ref{fig:0616c}(b), that the ground state energy (defined as the sum of all quasiparticle energies over occupied bands) becomes lower for $p_x - ip_y$ pairing than for $p_x + ip_y$ as $\mu$ is increased when the magnetic field is along the $+\hat z$ direction (as determined by the Peierls phase factors introduced below \eqref{eq:0428a}). This illustrates that, although the two possible chiralities of the order parameter couple to the magnetic field with opposite sign, determining which pairing state is energetically favorable is rather subtle, with the result depending on nonuniversal parameters that influence the detailed structure of the quasiparticle bands. (Note, however, that we have not determined the value of $\Delta$ from a fully self-consistent calculation, which should be performed in order to make a more definitive comparison between different candidate order parameters.)

Writing the Hamiltonian as $\mathcal{H}(\bk) = \sum_{i=1}^3 h_i (\bk) \sigma_i$, one finds that the Majorana bands have nontrivial Berry curvature \cite{kraus11}:
\be
\label{eq:0505b}
\Omega_\bk = \mp \frac{1}{|\mathbf{h}(\bk)|^3} \left( \mathbf{h}(\bk) \cdot
	\frac{\partial \mathbf{h}(\bk)}{\partial k_x}
	\times \frac{\partial \mathbf{h}(\bk)}{\partial k_y} \right),
\ee
with the upper (lower) sign corresponding to the higher- (lower-) energy band. 
Integrating the Berry curvature over momentum gives the Chern number of each Majorana band:
\be
\label{eq:0505b}
\nu = \mp \mathrm{sgn} (t_1),
\ee
where again the upper (lower) sign corresponds to the higher- (lower-) energy band. 

Finally, although the focus so far has been on the square vortex lattice, we note that the effective Hamiltonian \eqref{eq:0505a} can also describe non-square vortex lattices with $L_x \neq L_y$ if the next-nearest-neighbor hopping terms are allowed to be different along different directions, i.e.~the $\sigma_3$ term is replaced by $t_1 \sin(k_x L_x) - t_1' \sin(k_y L_y)$. Such anisotropic vortex lattices do not in general have the Dirac cones that appear in the square lattice, even when further-neighbor hopping is tuned to zero. The Majorana bands still have Chern number $\pm 1$. In the special case of a triangular vortex lattice \footnote{Note that a triangular vortex lattice can only be approximated in our theory due to the fact that $L_x / L_y$ must be a rational number.}, where $L_x / L_y = \sqrt{3}$, one has $(t_1, t_1') \to (0, t_0)$, and the Chern numbers of the Majorana bands are simply determined by the sign of $t_0$, as shown in previous work \cite{kraus11}.

\section{Edge states and thermal conductivity}
\begin{figure}
\centering
\includegraphics[width=0.48\textwidth]{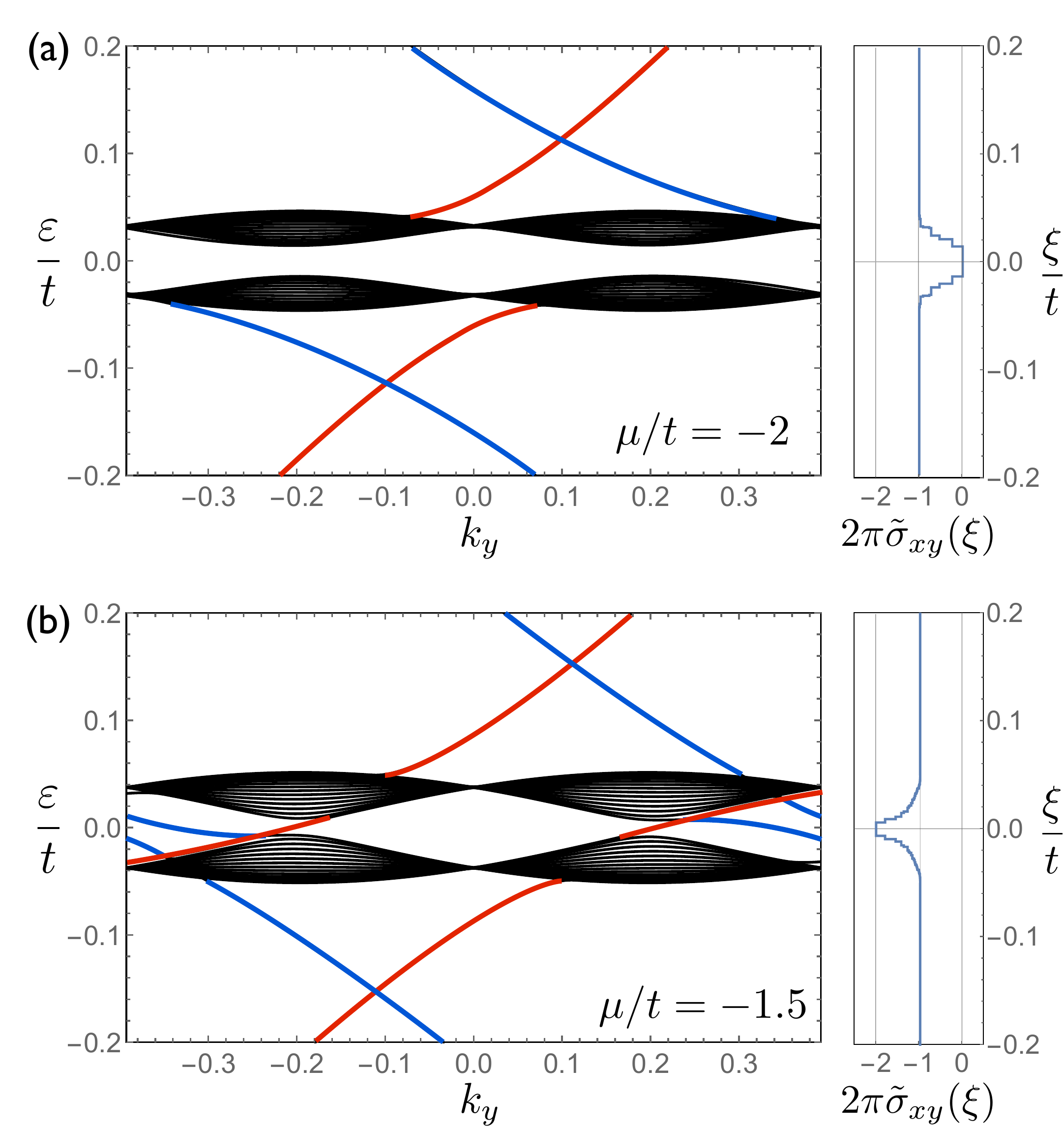}
\caption{Energy eigenvalues for a chiral p-wave superconductor (with $p_x - i p_y$ pairing and magnetic length $L_x = L_y = 8$) in the vortex state on a cylinder with periodic boundary conditions along the $y$-direction and open boundary conditions along the $x$-direction. The right panels show the corresponding integrated Berry curvature summed over occupied bands. (a) With $\mu / t = -2$, the gap between Majorana bands is topologically trivial due to cancellation of the Berry curvature of the Majorana bands with that of the superconducting condensate, so that no edge modes are present. At energies above and below the Majorana bands, however, states localized at the right (red) and left (blue) edges are present due to the nontrivial topology of the background superconducting condensate. (b) With $\mu / t = -1.5$, the gap between Majorana bands is topologically nontrivial due to constructive addition of the Berry curvature of the Majorana bands with that of the condensate, so that two chiral edge modes are present .
\label{fig:0616b}}
\end{figure}

When the topologically nontrivial Majorana bands are considered within the background of the chiral p-wave condensate, two possibilities arise: either the Chern numbers of the Majorana bands add constructively with those of the condensate, giving overall Chern number $\nu = \pm 2$, or they add destructively, giving $\nu = 0$. A phase diagram showing these cases is shown in Figure \ref{fig:0616c}(c).
The edge states for the two cases are shown in Figure \ref{fig:0616b}, from which it can be seen that the number of edge modes corresponds to the Chern number in each case. In particular, in both cases there is a single chiral edge mode in the large energy gaps above and below the Majorana bands, reflecting the chiral nature of the superconducting condensate. (These gaps could be accessed by shifting the Majorana bands with a Zeeman term, which is not included in the figure.) 
Figure \ref{fig:0616b}(a) shows a case in which the Chern number arising from the Majorana bands cancels with that of the background condensate, so that there are no edge modes near energy $\varepsilon = 0$. In contrast, Figure \ref{fig:0616b}(b) shows a case in which there are two chiral edge modes near $\varepsilon=0$, with one arising from the superconducting condensate and the other arising from the topological Majorana bands.

\begin{figure}
\centering
\includegraphics[width=0.48\textwidth]{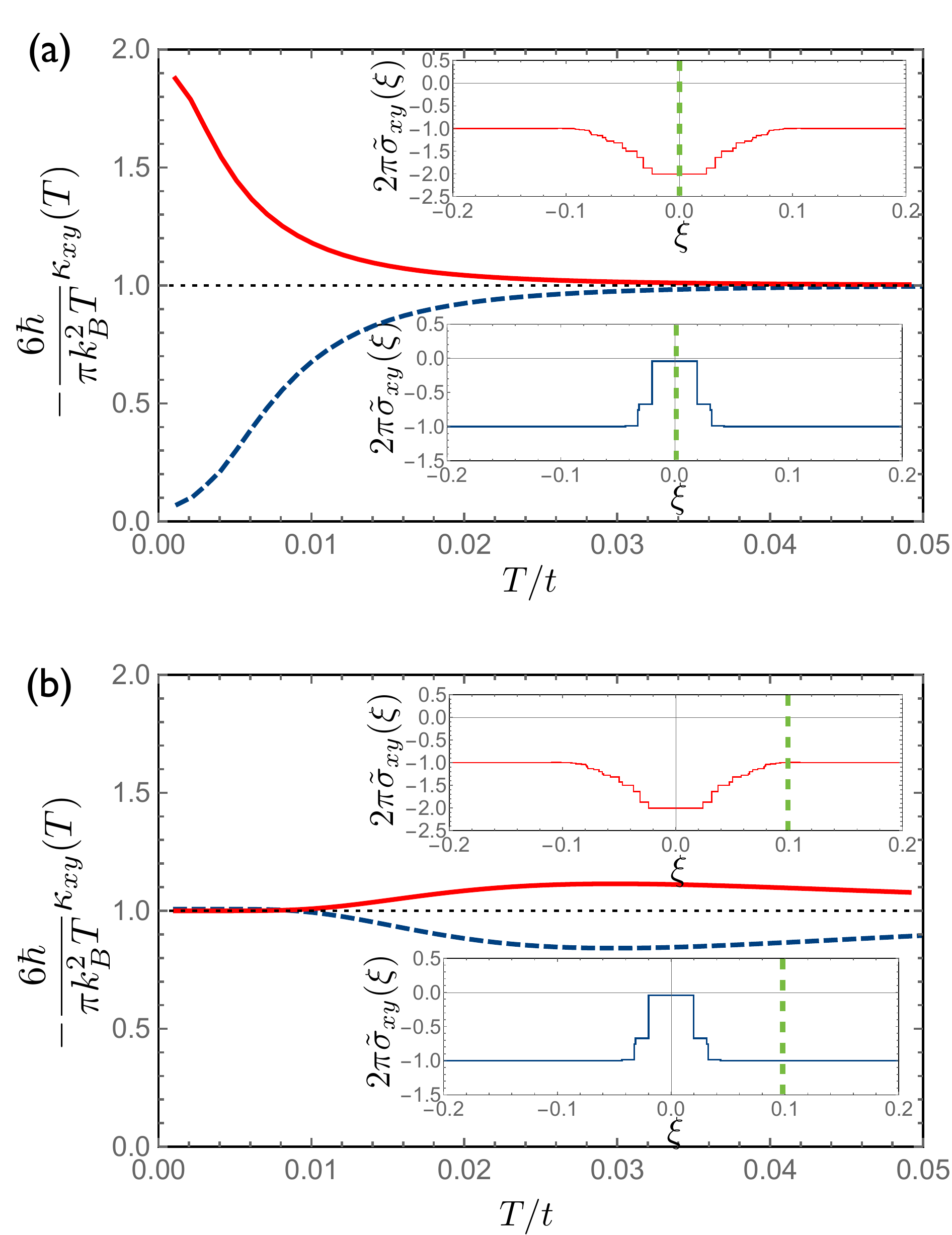}
\caption{Thermal Hall conductivity divided by temperature for $\Delta/t = 0.5$, $L_{x,y} = 8$, and $p_x - i p_y$ pairing. The black dotted line corresponds to the quantized value for a system with Chern number $\nu=-1$. Solid red and dashed blue lines show the results for the cases where the gap around energy $\xi=0$ has Chern number $\nu = -2$ (at $\mu / t = -2$) and $\nu = 0$ (at $\mu / t = -1.5$), respectively (see phase diagram in Figure \ref{fig:0616c}(c)). Insets show the corresponding $\tilde \sigma_{xy} (\xi)$, which determine $\kappa_{xy} (T)$ via \eqref{eq:0609a}. (a) With no Zeeman coupling, the low-temperature behavior is determined by $\tilde \sigma_{xy} (\xi)$ near $\xi \sim 0$. (b) With Zeeman coupling $h_Z = 0.1 t$, the low-temperature behavior is determined instead by $\tilde \sigma_{xy} (\xi)$ near $\xi \sim h_Z$.
\label{fig:0619a}}
\end{figure}

Although these edge modes do not contribute to electrical transport due to the fact that they occur within a superconducting state, they do contribute a thermal current in the presence of a temperature gradient. The thermal Hall conductivity is given by \cite{smrcka77,vafek01b,cvetkovic15}
\be
\label{eq:0609a}
\kappa_{xy} = \frac{1}{\hbar T} \int_{-\infty}^\infty d\xi \xi^2
	\left( - \frac{\partial n_F (\xi)}{\partial \xi} \right)
	\tilde \sigma_{xy} (\xi ),
\ee
where $n_F (\xi) = 1/(e^{\xi/T}+1)$ is the Fermi occupation factor. The thermal Hall conductivity is thus determined by the integrated Berry curvature, convolved with a thermal factor that is peaked near $\xi = 0$.
This quantity is plotted in Figure \ref{fig:0619a}. 
In Figure \ref{fig:0619a}(a) we show the result without Zeeman shift ($h_Z = 0$), and in this case $\kappa_{xy}(T)/T$ is determined by the $\tilde\sigma_{xy}(\xi)$ near energy $\xi = 0$ at low temperature. At higher temperature, the thermal factor in \eqref{eq:0609a} broadens beyond the width of the Majorana bands, so that $\kappa_{xy}(T)/T$ is instead determined by $\tilde\sigma_{xy}(\xi)$ at energies above and below the Majorana bands. 
Figure \ref{fig:0619a}(b) shows $\kappa_{xy}(T)/T$ with Zeeman shift. As noted above, the effect of Zeeman coupling is simply to shift the overall energy of the quasiparticle bands, so that $\tilde\sigma_{xy}(\xi) \to \tilde\sigma_{xy}(\xi-h_Z)$, and $\kappa_{xy}(T)$ is determined via \eqref{eq:0609a} by the value of $\tilde\sigma_{xy}$ near $\xi \sim h_Z$. In this example, we choose the Zeeman energy $h_Z$ to be half of the cyclotron energy, as it is for free electrons: $h_Z = \hbar\omega_c/2 = 2\pi t /(L_x L_y)$, where we have expanded the tight-binding dispersion near its minimum to obtain the effective mass $m = \hbar^2 / (2t)$. In this case $\kappa_{xy}(T)/T$ is determined by $\tilde\sigma_{xy}(\xi\sim h_Z)$ at low temperatures, while it shows a modest change as the thermal factor in \eqref{eq:0609a} broadens at higher temperatures, increasing or decreasing to reflect the behavior of $\tilde\sigma_{xy}(\xi)$ near $\xi \sim 0$.

\section{Conclusion}

The results of our work may have relevance for the candidate topological superconductor Sr$_2$RuO$_4$, for which very clean samples are available and thermal conductivity measurements are feasible. Because it is a multiband superconductor, however, a more realistic lattice model should be used if quantitative comparison is to be made. Our results may also be relevant for the Moore-Read fractional quantum Hall state, which can be thought of as a gas of composite fermions with chiral p-wave pairing.


In conclusion, we have shown that the topological nature of the vortex state of a chiral p-wave superconductor at low energies is far from being obvious, generally requiring a complete microscopic calculation to determine reliably. In particular, the result depends on the constructive or destructive addition of Chern numbers associated with the Majorana bands and the superconducting condensate. More broadly, our work illustrates a way in which thermal conductivity can be a useful probe for obtaining information about the quantum wavefunction describing an interacting many-body system. Such a probe is especially useful in a superconductor, where electrical transport measurements may not yield useful information.

\section{Acknowledgements}
This work was supported by the NSF CAREER award
under Grant No.~DMR-0955561, NSF Cooperative
Agreement No.~DMR-1157490 and the State of Florida.

\bibliographystyle{apsrev}
\bibliography{refs}

\end{document}